\documentclass[a4paper,11pt]{article}
\usepackage{pos}

\title{Resummation of threshold double logarithms in inclusive production of
heavy quarkonium}
\ShortTitle{Resummation of threshold double logarithms in 
quarkonium production}

\author*[a]{Hee Sok Chung}
\author[b]{U-Rae Kim}
\author[a]{Jungil Lee}

\affiliation[a]{Department of Physics, Korea University, Seoul 02841, Korea}

\affiliation[b]{Department of Physics, Korea Military Academy, Seoul 01805, Korea}

\emailAdd{neville@korea.ac.kr}
\emailAdd{kim87@kma.ac.kr}
\emailAdd{jungil@korea.ac.kr}

\abstract{
We resum threshold double logarithms in inclusive production of heavy
quarkonium that arise from singularities near the boundary of phase space. 
This resolves the catastrophic failure in the conventional approach based on
fixed-order perturbation theory calculations in nonrelativistic QCD, where
quarkonium cross sections at large transverse momentum can turn negative. 
We identify the root cause of this negative cross section problem as the
appearance of threshold logarithms in radiative corrections, and resum them to
all orders in perturbation theory at the leading double logarithmic level. 
We find that resummation of threshold logarithms is imperative for describing 
measured $J/\psi$ production rates at large transverse momentum. 
}

\FullConference{The XVIth Quark Confinement and the Hadron Spectrum Conference (QCHSC24)\\
 19-24 August, 2024\\
 Cairns Convention Centre, Cairns, Queensland, Australia\\}

\begin{document}
\maketitle

\section{Introduction}

Heavy quarkonium production phenomenology faced a serious challenge when 
in 2019, the
ATLAS Collaboration released a measurement of prompt $J/\psi$ cross section
at unprecedentedly large values of transverse momentum reaching up to 360~GeV, 
based on their 13~TeV data~\cite{ATLAS:2019ilf}. Soon, some theorists noticed 
that computing the prompt $J/\psi$ cross section at such large values of $p_T$
is problematic when using 
fixed-order calculations at next-to-leading order (NLO) accuracy in 
the nonrelativistic QCD (NRQCD) factorization formalism~\cite{Bodwin:1994jh}.
While a sizable amount of prompt $J/\psi$ is produced
through decays of $\chi_{cJ}$, these cross sections computed in NLO NRQCD
were turning negative at large $p_T$, making it impossible to make a
trustworthy prediction. It was also found that if one wants to describe the
$J/\psi$ and $\eta_c$ hadroproduction rates simultaneously, the direct $J/\psi$
cross section was turning negative at values of $p_T$ dangerously close to the
maximum $p_T$ range of the ATLAS data, which made the prediction unreliable. 
We refer readers to refs.~\cite{Chung:2018lyq, Chung:2022uih} on the current
status of heavy quarkonium production phenomenology based on NLO NRQCD. 

This negative cross section problem remained unresolved for several years which
triggered a public discussion in 2022~\cite{hsctalk}. 
By this time theorists began to realize that this problem had something to do
with logarithms associated with the boundary of phase space~\cite{gtb}. 
These kinds
of logarithms are named {\it threshold logarithms} and the only cure for
problems caused by them is to resum them to all orders in perturbation
theory~\cite{Ivanov:1985np, Korchemsky:1985ts, Korchemsky:1986fj, 
Sterman:1986aj, Korchemsky:1988si, Catani:1989ne, Catani:1990rp, 
Korchemsky:1993uz}. 
After some preliminary studies~\cite{Chen:2021hzo, Chung:2023ext,
Chen:2023gsu}, a complete resummation of threshold double logarithms at leading
logarithmic level was finally accomplished in ref.~\cite{Chung:2024jfk},
leading to positive-definite cross section predictions based on first
principles that well describe the ATLAS data~\cite{ATLAS:2023qnh}. 

In this proceeding based on our recent work in ref.~\cite{Chung:2024jfk}, 
we summarize the
main results of the resummation of threshold double logarithms in inclusive 
production of $J/\psi$, $\psi(2S)$, and $\chi_{cJ}$ in hadron colliders, which
leads to the cure of the negative cross section problem. 
We introduce the NRQCD factorization formalism in
Sec.~\ref{sec:NRQCDfac}, and show the appearance of threshold logarithms in
Sec.~\ref{sec:sdcs}. We then describe the resummation of threshold logarithms
in Sec.~\ref{sec:softfac}, and show numerical results in Sec.~\ref{sec:numer}.
We conclude in Sec.~\ref{sec:summary}. 

%==============================================================================
\section{NRQCD factorization approach to heavy quarkonium production}
\label{sec:NRQCDfac}
%==============================================================================

The large-$p_T$ cross section of a heavy quarkonium ${\cal Q}$ is given
in the NRQCD factorization formalism by~\cite{Bodwin:1994jh} 
%---------------
\begin{equation}
\label{eq:NRQCDfac}
%---------------
\sigma_{{\cal Q}} = \sum_{\cal N} 
\sigma_{Q \bar Q ({\cal N})} \langle {\cal O}^{\cal Q}({\cal N}) \rangle, 
%---------------
\end{equation}
%---------------
where $\sigma_{Q \bar Q ({\cal N})}$ is the short-distance coefficient (SDC)
for production of a heavy quark ($Q$) and a heavy 
antiquark ($\bar Q$) pair in the state ${\cal N}$, 
and $\langle {\cal O}^{\cal Q}({\cal N}) \rangle$ is
the long-distance matrix element (LDME) that corresponds to the probability 
for a $Q \bar Q$ in the state $\cal N$ to evolve into a quarkonium ${\cal Q}$, 
for which we adopt the normalization in ref.~\cite{Bodwin:1994jh}. 
The sum over $\cal N$ is usually truncated at a desired order in the
nonrelativistic expansion. For ${\cal Q} = \chi_{cJ}$, the 
$^3P_J^{[1]}$ and $^3S_1^{[8]}$ channels appear at leading order in the
expansion; for ${\cal Q} = J/\psi$ or $\psi(2S)$, dominant
contributions come from $^3S_1^{[1]}$, $^3S_1^{[8]}$, $^1S_0^{[8]}$, and
$^3P_J^{[8]}$ channels\footnote{This follows from that in hadroproduction, 
the SDC of the  $^3S_1^{[1]}$ channel is strongly suppressed
compared to the color-octet channels, even though the color-octet contributions
are subleading in the
nonrelativistic expansion. In cases where the color-singlet contribution is not
suppressed, relativistic corrections to the color-singlet channel may be more
important than the color-octet contributions. 
}. 

\begin{figure}[t]
\centering
\includegraphics[width=.99\columnwidth]{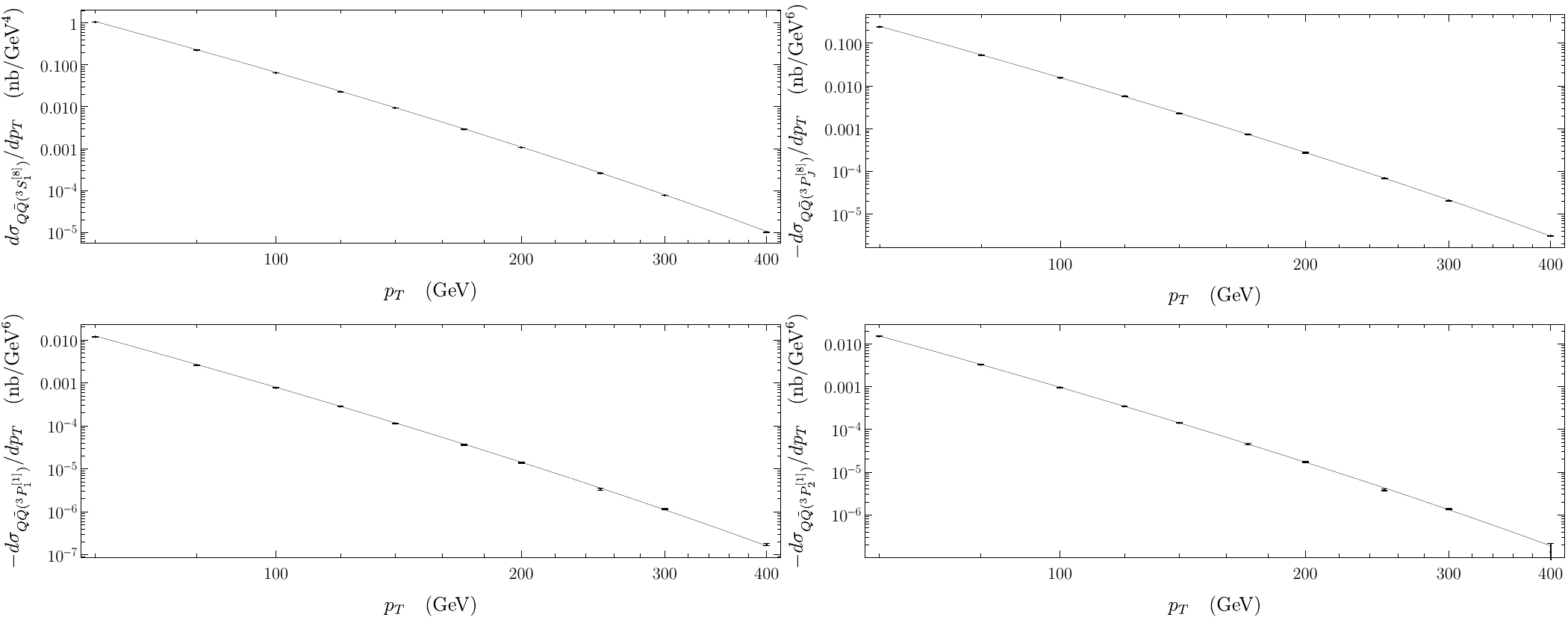}%
\caption{\label{fig:SDCs} 
Comparison of $\sigma_{Q \bar Q ({\cal N})}^{\rm LP}$ (lines) 
and $\sigma_{Q \bar Q ({\cal N})}$ (points) computed at $\alpha_s^4$ accuracy
for ${\cal N} = {}^3S_1^{[8]}$, $^3P_J^{[8]}$, $^3P_1^{[1]}$, and $^3P_2^{[1]}$ 
in $pp$ collisions at $\sqrt{s}=7$~TeV, $|y|<1.2$.
Error bars correspond to numerical uncertainties in 
$\sigma_{Q \bar Q ({\cal N})}$. 
For $P$-wave channels, negative values are shown because the cross sections
turn negative at large $p_T$. 
}
\end{figure}

In the large-$p_T$ region, the SDCs are given by
leading-power (LP) fragmentation~\cite{Collins:1981uw, Nayak:2005rt}
%---------------
\begin{equation}
\label{eq:LPfrag}
%---------------
\sigma_{Q \bar Q ({\cal N})}^{\rm LP} = \sum_{i=g,q,\bar{q}} 
\int_0^1 dz \, \hat{\sigma}_{i(K)} D_{i \to Q \bar Q ({\cal N})} (z), 
%---------------
\end{equation}
%---------------
where 
$\hat{\sigma}_{i (K)}$ is the cross section of a parton $i = g$, 
$q$, $\bar{q}$ with momentum $K$, 
$D_{i \to Q \bar Q ({\cal N})} (z)$ is the fragmentation 
function (FF) for fragmentation of $i$ into $Q \bar Q ({\cal N})$, 
and 
$z = P^+/K^+$ is the fraction of the $Q \bar Q$ momentum $P$ compared to $K$ 
projected onto a $+$ direction, defined through a lightlike vector $n$ with 
$K^+ = n \cdot K$. 
The correction to Eq.~(\ref{eq:LPfrag}) is suppressed by $m^2/p_T^2$, 
with $m$ the heavy quark mass, and can be computed from 
next-to-leading power (NLP) fragmentation~\cite{Kang:2011zza,
Kang:2014tta, Ma:2014svb}. 
The equivalence of Eqs.~(\ref{eq:NRQCDfac}) and (\ref{eq:LPfrag}) at large
$p_T$ is ensured by the collinear factorization theorem of
QCD~\cite{Collins:1981uw, Nayak:2005rt}, 
which holds independently of the final state. 
It is also possible to verify this numerically order by order in perturbation 
theory, as was done in ref.~\cite{Bodwin:2015iua}. 
In Fig.~\ref{fig:SDCs} we compare the order-$\alpha_s^4$ (NLO) calculations
of $\sigma_{Q \bar Q ({\cal N})}$, 
which was computed in ref.~\cite{Bodwin:2015iua}, 
and $\sigma_{Q \bar Q ({\cal N})}^{\rm LP}$ 
for ${\cal N} = {}^3S_1^{[8]}$, $^3P_J^{[8]}$, $^3P_1^{[1]}$, and $^3P_2^{[1]}$ 
for $p_T$ up to 400~GeV. 
Results for other channels, including polarized ones, can be found in
ref.~\cite{Bodwin:2015iua}. 
In every case we see excellent agreement, up to
uncertainties due to numerical integration used in computing 
$\sigma_{Q \bar Q ({\cal N})}$. 
Despite dubious claims in the literature, the validity of
Eq.~(\ref{eq:LPfrag}) has been shown to be robust and reliable for heavy
quarkonium production at large $p_T$.

%%%%%%%%%%%%%%%%%%%%%%%%%%%%%%%%%%%%%%%%%%%%%%%%%%%%%%%%%%%%%%%%%%%%%%%%%%%%%
%%%%%%%%%%%%%%%%%%%%%%%%%%%%%%%%%%%%%%%%%%%%%%%%%%%%%%%%%%%%%%%%%%%%%%%%%%%%%

%==============================================================================
\section{Threshold logarithms and the negative cross section problem}
\label{sec:sdcs}
%==============================================================================

The gluon FFs for the $^3S_1^{[8]}$ and the $P$-wave channels at the lowest
nonvanishing orders in $\alpha_s$ are known to involve distributions that are
singular at $z=1$, whereas all other FFs are regular functions in $z$. 
To order-$\alpha_s^2$ accuracy, the singular FFs are given
by~\cite{Braaten:1996rp, Braaten:2000pc, Lee:2005jw, Bodwin:2012xc, Ma:2013yla}
%---------------
\begin{align}
%---------------
D_{g \to Q \bar Q(^3S_1^{[8]})} (z) &= 
\frac{\pi \alpha_s(\mu_R)}{24 m^3} 
\bigg\{ \delta (1-z) + \frac{\alpha_s}{\pi} \bigg[
A(\mu_R) \delta (1-z) + 
\left( \log\frac{\mu_F}{2 m} - \frac{1}{2} \right) P_{gg} (z)
\nonumber \\ & \hspace{2ex} 
+ \frac{3 (1-z)}{z} 
+ 6 (2-z+z^2) \log(1-z) - 
\frac{2 C_A}{z} \left( \frac{\log(1-z)}{1-z} \right)_+ 
\bigg] \bigg\} + O(\alpha_s^3),
\\
D_{g \to Q \bar Q({}^3P_J^{[8]})} (z) &= 
\frac{8 \alpha_s^2}{9 (N_c^2-1) m^5} 
\frac{N_c^2-4}{4 N_c} 
\bigg[  \delta (1-z) \left(\frac{1}{6}
-  \log \frac{\mu_\Lambda}{2 m} \right)
\nonumber \\ & \hspace{5ex}
+ \left( \frac{1}{1-z} \right)_+ 
+ \frac{13-7 z}{4} \log (1-z) 
- \frac{(1-2 z) (8-5 z)}{8} \bigg]+O(\alpha_s^3), 
%---------------
\end{align}
%---------------
with $A(\mu_R) = \frac{\beta_0}{2} \left( \log \frac{\mu_R}{2 m_c} +
\frac{13}{6} \right) + \frac{2}{3} - \frac{\pi^2}{2} + 8 \log 2$,
$\beta_0 = \frac{11}{3} N_c+\frac{2}{3} n_f$,  
and $P_{gg} (z)$ the gluon splitting function. 
The $\mu_\Lambda$ is the NRQCD factorization scale in the $\overline{\rm
MS}$ scheme, which is often taken to be the heavy quark mass. 
The FFs for the color-singlet $P$-wave states have similar forms as the 
${}^3P_J^{[8]}$ FF. 
Note that the $P$-wave FFs vanish at order $\alpha_s$. 
The severity of the singularities of the Dirac delta and the plus functions 
can be quantified by taking the Mellin transform 
%---------------
\begin{equation}
\label{eq:Mellin}
%---------------
\tilde{D}_{i \to Q \bar Q ({\cal N})} (N) = \int_0^1 dz \, z^{N-1} 
D_{i \to Q \bar Q ({\cal N})} (z), 
%---------------
\end{equation}
%---------------
and examining the behavior at $N \to \infty$. 
While the Mellin transform of $\delta (1-z)$ is $1$, the 
$\left( \frac{\log(1-z)}{1-z} \right)_+$ term in the $^3S_1^{[8]}$ FF diverges
like $\log^2 N$ in Mellin space. 
The strongest divergence in the $^3P_J^{[8]}$ FF comes from
the $\left( \frac{1}{1-z} \right)_+$ term, whose Mellin transform diverges like
$\log N$, while the order-$\alpha_s^3$ correction is known to involve a term
proportional to $\left( \frac{\log^2(1-z)}{1-z} \right)_+$, which diverges like
$\log^3 N$~\cite{Zhang:2020atv}. 
These terms in the radiative corrections are called threshold
double logarithms, as they involve the boundary of phase space ($z=1$), and are
proportional to $\log^2 N$ times the leading-order FF. Corrections of higher
orders in $\alpha_s$ will involve corrections proportional to $(\alpha_s \log^2
N)^n$, which will jeopardize the convergence of perturbation theory. 

The effect of the $z=1$ singularities in the FFs to the cross section 
is amplified with increase in $p_T$,
as the steep rise in $z$ of the parton cross sections 
$\hat{\sigma}_{i(K)}$ becomes increasingly severe. 
Because of this, 
the threshold double logarithms modify the $p_T$ shapes of the cross sections.
Note that since the $\hat{\sigma}_{i(K)}$ begin at order $\alpha_s^2$, and the
FFs begin at order $\alpha_s$, the fixed-order NLO calculations contain
contributions from FFs only up to order-$\alpha_s^2$ accuracy. Hence, in a
fixed-order calculation, only the $^3S_1^{[8]}$ channel contains the effect of
the threshold double logarithm at NLO, while the $P$-wave channels do not until
NNLO. As a result, at NLO the ${}^3S_1^{[8]}$ cross section falls off faster
than the ones in the $P$-wave channels, as have been shown in refs.~\cite{
Chung:2023ext, Chen:2023gsu}. Consequently, when the quarkonium cross section
involves a cancellation between the ${}^3S_1^{[8]}$ and a $P$-wave channel
contribution, the cross section will eventually turn negative at some large
value of $p_T$. 
{\it This is the root cause of the negative cross section
problem~\cite{hsctalk}.}

It has been shown that depending on the choice of LDMEs, the sign change in the
$J/\psi$, $\psi(2S)$, or $\chi_{cJ}$ cross sections can happen below
$p_T=360$~GeV, leading to a failure to describe ATLAS data~\cite{ATLAS:2019ilf,
ATLAS:2023qnh}. To make matters worse, the polarized cross sections can even
turn negative at much lower values of $p_T$ than the polarization-summed one.
Although it may be possible to defer this catastrophe until a larger value of
$p_T$ by adjusting the LDMEs, the prediction will still be unreliable if the
sign change happens dangerously close to the $p_T$ range of interest, as the
theory prediction would be distorted by an arbitrary truncation of large
logarithmic corrections. 
The LDMEs would also be skewed by improper treatment of the threshold
logarithms, which can deteriorate the universality of LDMEs and the predictive
power of the effective field theory formalism. 
It is important to note that because threshold logarithms appear in different
channels at different orders in $\alpha_s$, a truncation of the perturbation
series at an arbitrarily chosen accuracy will always lead to inconsistencies. 
Therefore, the only proper way to resolve the negative cross section problem 
is to resum threshold logarithms to all orders.

%==============================================================================
\section{Soft factorization and resummation}
\label{sec:softfac}
%==============================================================================

The resummation of threshold logarithms can be accomplished by first obtaining
an approximate expression for the FFs that reproduce the $z \to 1$
singularities to all orders in $\alpha_s$. This can be done by application of
the Grammer-Yennie approximation (soft approximation) for gluon attachments to
quark lines~\cite{Grammer:1973db, Collins:1981uk, Nayak:2005rt}. 
We list here the results obtained in~\cite{Chung:2024jfk}:
%---------------
\begin{subequations}
\begin{align}
\label{eq:Dsoft}
%---------------
D_{g \to Q \bar Q(^3S_1^{[8]})}^{\rm soft} (z)
&= \frac{C_{\rm frag} (d-2) g^2}{4 m^3 (d-1) (N_c^2-1)} S_{^3S_1^{[8]}} (z), 
\\
D_{g \to Q \bar Q(^3P_J^{[8]})}^{\rm soft} (z)
&= - \frac{C_{\rm frag} (d-2) g^4
}{4 m^3 (d-1)^2 (N_c^2-1)} 
S_{^3P^{[8]}} (z) 
,
\\
D_{g \to Q \bar Q(^3P_J^{[1]})}^{\rm soft} (z)
&= - \frac{C_{\rm frag} (d-2) g^4}{4 N_c^2 m^3 (d-1)^2} 
\frac{9}{(2 J+1)}
%\nonumber \\ & \quad \times \!
\left[ c_J S_{^3P^{[1]}} (z) 
+ c_J^{TT} S^{TT}_{^3P^{[1]}} (z) \right], 
%---------------
\end{align}
\end{subequations}
%---------------
where $C_{\rm frag} = z^{d-3} K^+/[2 \pi (N_c^2-1) (d-2)]$, 
$d=4-2 \epsilon$ is the number of spacetime dimensions,  
$g$ is the strong coupling, 
$c_0 = (d-1)^{-2}$, $c_1 = (d-2)/[2 (d-1)]$, $c_2 = (d-2) (d+1)/[2
(d-1)^2]$, $c_0^{TT} = [(d-1) (d-2)]^{-1}$, $c_1^{TT} = -[2 (d-2)]^{-1}$, 
$c_2^{TT} = (d-3)/[2 (d-1) (d-2)]$, 
and the soft functions $S_{\cal N} (z)$ are defined by 
%---------------
\begin{subequations}
\begin{align}
\label{eq:soft}
%---------------
S_{^3S_1^{[8]}} (z) &=
\langle 0 | [ \Phi_p^{ca} \Phi_n^{ba} ]^\dag 
2 \pi \delta(n \cdot \hat p - (1-z) P^+)
\Phi_p^{cd} \Phi_n^{bd} | 0 \rangle,
\\
S_{^3P^{[8]}} (z) &=
\langle 0 | [ {\cal W}^{yx}_{\alpha} ]^\dag
%\delta^+ _z
2 \pi \delta(n \cdot \hat p - (1-z) P^+)
{\cal W}_\beta^{y x} | 0 \rangle g^{\alpha \beta},
\\
S_{^3P^{[1]}} (z) &=
\langle 0 | [ {\cal \bar W}^{b}_{\alpha} ]^\dag
%\delta^+ _z
2 \pi \delta(n \cdot \hat p - (1-z) P^+)
{\cal \bar W}_\beta^{b} | 0 \rangle g^{\alpha \beta},
\\
S_{^3P^{[1]}}^{TT} (z) &=
\langle 0 | [ {\cal \bar W}^{b}_{\alpha}]^\dag
%\delta^+ _z
2 \pi \delta(n \cdot \hat p - (1-z) P^+)
{\cal \bar W}_\beta^{b} | 0 \rangle 
%\nonumber \\ & \quad \times 
\left[ \frac{p^2 n^\alpha n^{\beta}}{(n \cdot p)^2} 
+ \frac{g^{\alpha \beta}}{d-1} \right], 
%---------------
\end{align}
\end{subequations}
%---------------
where $\hat p$ is an operator that reads off the momentum of the operator to
the right, $\Phi_k (y,x)$ is an adjoint Wilson line along a vector $k$ from
point $xk$ to point $yk$, $\Phi_k \equiv \Phi_k (\infty,0)$, 
${\cal W}^{yx}_\beta = \int_0^\infty d \lambda \, \lambda
\Phi_p^{yc} (\infty, \lambda) p^\mu G^b_{\mu \beta} (p \lambda) d^{bcd} 
\Phi_p^{da} (\lambda,0) \Phi_n^{xa}$,
${\cal \bar W}^{b}_\beta = \int_0^\infty d \lambda \, \lambda
p^\mu G^d_{\mu \beta} (p \lambda) 
\Phi_p^{da} (\lambda,0) \Phi_n^{ba}$, 
with $G_{\mu \nu}$ the field-strength tensor, 
and $p = P/2$. 
The operators on the right and left of the Dirac delta function are time and
anti-time ordered, respectively, which is implicit in these expressions. 

It is worth noting that the soft function $S_{^3S_1^{[8]}} (z)$ is given by a
Wilson loop; in this case, the double logarithmic corrections this soft
function is given in terms of the cusp anomalous dimension. In the case of the
Wilson loop in the fundamental representation, the anomalous dimension is known
to three loops~\cite{Bruser:2019yjk}. At one loop, the calculation in the
fundamental representation can be translated to the adjoint representation 
by making the replacement $C_F \to C_A$. Although the threshold logarithms in
the $^3S_1^{[8]}$ channel has been investigated in ref.~\cite{Chen:2021hzo},
the universal form of the soft function $S_{^3S_1^{[8]}} (z)$ has been first
obtained in~\cite{Chung:2024jfk}.  
The soft function $S_{^3P^{[1]}}(z)$ was first derived in the shape-function
calculation of $\chi_{cJ}$ production rates in ref.~\cite{Chung:2023ext}. 
The $S_{^3P^{[8]}} (z)$ and $S_{^3P^{[1]}}^{TT} (z)$ are new in 
ref.~\cite{Chung:2024jfk}. 

The threshold double logarithms arise from the double UV poles of the loop
corrections to the soft functions. By explicit calculation we can show that 
$S_{^3P^{[1]}}^{TT} (z)$ does not produce double logarithms at NLO, 
although it does produce single logarithms and is necessary for computing the
$J$-dependent single logarithms in the $^3P_J^{[1]}$ FFs. 
From the double UV poles in the NLO corrections to the soft functions, we 
obtain expressions for the double logarithms in ${\cal S}_{\cal N}(z)$ 
written in Mellin space as 
$J_{\cal N}^{N} \tilde {\cal S}_{\cal N}^{\rm LO} (N)$, 
with $J_{^3S_1^{[8]}}^N = \frac{\alpha_s C_A}{\pi} 
 \int_0^1 dz \, z^{N-1} \left[ \frac{-2 \log(1-z)}{1-z} \right]_+$ 
and $J_{^3P^{[8]}}^ N= J_{^3P^{[1]}}^N = \frac{4}{3} J_{^3S_1^{[8]}}^N$.
Because the double UV poles only arise from planar diagrams, they can be
immediately exponentiated. 
From this we obtain the threshold-resummed expression for the FFs given by 
%---------------
\begin{align}
\label{eq:Dresum}
%---------------
\tilde D_{g \to Q \bar Q ({\cal N})}^{\rm resum} (N) 
&=
\exp\left[ J_{\cal N}^N \right]
\times 
\bigg(
\tilde D_{g \to Q \bar Q ({\cal N})}^{\rm FO} (N) 
- J_{\cal N}^N 
\tilde D_{g \to Q \bar Q ({\cal N})}^{\rm LO} (N)
\bigg), 
%---------------
\end{align}
%---------------
where the second term in the parenthesis subtracts the double logarithmic
correction in $\tilde D_{g \to Q \bar Q({\cal N})}^{\rm FO} (N)$ 
at NLO accuracy to avoid double counting. 
Note that because $\exp\left[ J_{\cal N}^N \right]$ vanishes faster than any
power of $N$, which makes the inverse Mellin transform rapidly convergent, 
the resummed FF is a regular function in $z$ that vanishes at $z=1$.

%==============================================================================
\section{Numerical results}
\label{sec:numer}
%==============================================================================

%%%%%%%%%%%%%%%%%%%%%%%%%%%%%%%%%%%%%%%%%%%%%%%%%%%%%%%%%%%%%%%%%%%%%%%%%%%%%
\begin{figure}[t]
\centering
\includegraphics[width=.45\columnwidth]{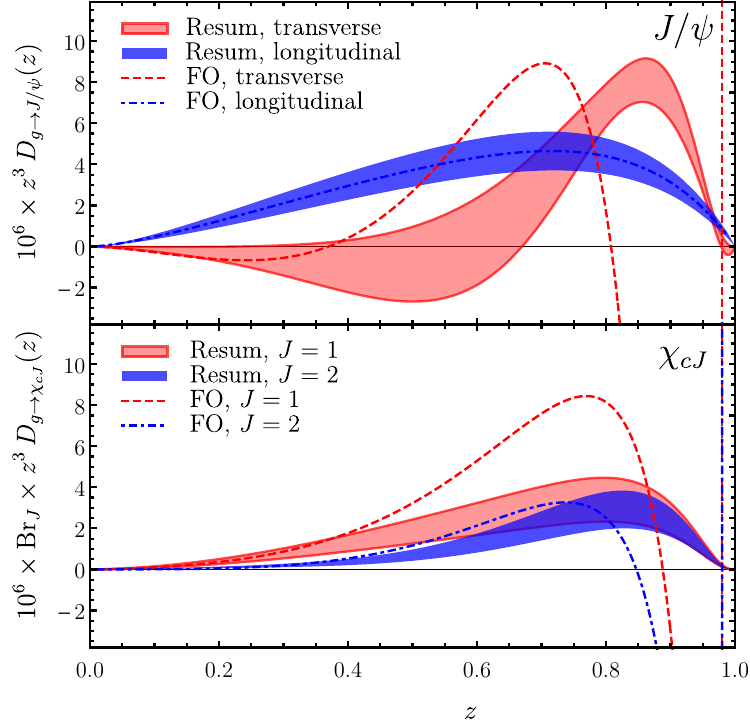}%
\caption{\label{fig:FFs} 
Gluon FFs with resummed threshold double logarithms 
times $z^3$ for production of $J/\psi$ (top) and $\chi_{cJ}$ (bottom) 
for $J=1$ and 2. Central values of FO results are also shown for comparison. 
${\rm Br}_J \equiv {\rm Br}_{\chi_{cJ} \to J/\psi + \gamma}$ is the branching
fraction for decays of $\chi_{cJ}$ into $J/\psi+\gamma$. 
Taken from ref.~\cite{Chung:2024jfk}. 
}
\end{figure}
%%%%%%%%%%%%%%%%%%%%%%%%%%%%%%%%%%%%%%%%%%%%%%%%%%%%%%%%%%%%%%%%%%%%%%%%%%%%%

We first show in Fig.~\ref{fig:FFs}
the resummed FFs for $J/\psi$ and $\chi_{cJ}$ defined by 
$D_{g \to {\cal Q}} (z) = \sum_{\cal N} 
D_{g \to Q \bar Q ({\cal N})} (z)
\langle {\cal O}^{\cal Q} ({\cal N}) \rangle$,  
where the sum runs over 
${\cal N} = {}^3S_1^{[1]}$, $^3S_1^{[8]}$, $^1S_0^{[8]}$,
and $^3P_J^{[8]}$ for ${\cal Q} = J/\psi$ or $\psi(2S)$, 
and ${\cal N} = {}^3P_J^{[1]}$ and $^3S_1^{[8]}$ for ${\cal Q} = \chi_{cJ}$.
Aside from threshold resummation, we also resum the
Dokshitzer-Gribov-Lipatov-Altarelli-Parisi (DGLAP)
logarithms~\cite{Gribov:1972ri, Lipatov:1974qm, Dokshitzer:1977sg,
Altarelli:1977zs} by evolving the FFs from the $\overline{\rm MS}$ scale 
3~GeV to 50~GeV. 
We use the LDMEs from ref.~\cite{Brambilla:2022ayc}, except we reduce the
central value of the $^3P^{[8]}$ LDME by 10\% in order to compensate for the
10\% enhancement of the resummed $^3P^{[8]}$ SDC compared to the $^3S_1^{[8]}$
one. The specific values of the LDMEs used in this calculation are listed in
ref.~\cite{Chung:2024jfk}.
We display the gluon FFs for transversely and longitudinally polarized $J/\psi$ 
in Fig.~\ref{fig:FFs}, which are, unlike the FO FFs, are positive or at least
consistent with zero within uncertainties for all $0 < z < 1$. This ensures
the positivity of $J/\psi$ production rates and resolves the negative cross
section problem. On the other hand, since the FO FFs change sign rapidly near 
$z=1$, the cross section can become negative due to the $z$ dependence of the
parton cross sections $\hat{\sigma}_{i(K)}$. 
The same is true for $\chi_{cJ}$ FFs; especially in this case, 
no choice of LDMEs will be able ensure positivity if threshold
resummation is not taken into account, because the FO FFs for both the 
$^3S_1^{[8]}$ and $^3P_J^{[1]}$ channels change sign rapidly near $z=1$.

%%%%%%%%%%%%%%%%%%%%%%%%%%%%%%%%%%%%%%%%%%%%%%%%%%%%%%%%%%%%%%%%%%%%%%%%%%%%%
\begin{figure}[t]
\centering
\includegraphics[width=.5\columnwidth]{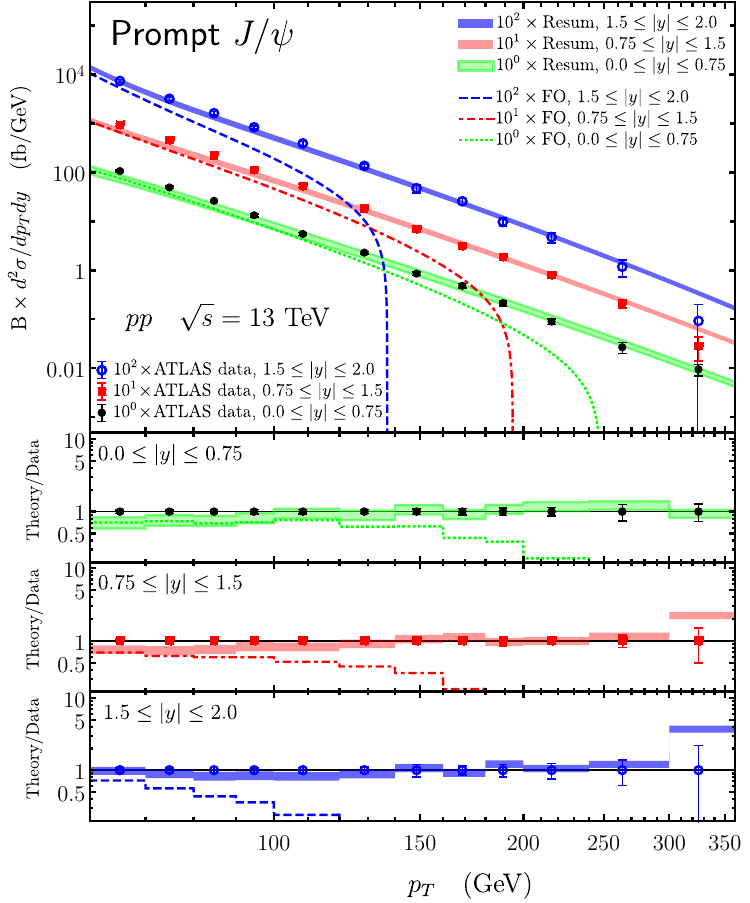}%
\caption{\label{fig:ATLAS}
Prompt $J/\psi$ production rates from $pp$ collisions at $\sqrt{s}=13$~TeV 
computed from resummed SDCs compared to ATLAS data. 
Central values of FO NLO results are shown for comparison. 
${\rm B} \equiv {\rm Br}_{J/\psi \to \mu^+ \mu^-}$ is the
$J/\psi$ dimuon branching fraction. 
Taken from ref.~\cite{Chung:2024jfk}. 
}
\end{figure}
%%%%%%%%%%%%%%%%%%%%%%%%%%%%%%%%%%%%%%%%%%%%%%%%%%%%%%%%%%%%%%%%%%%%%%%%%%%%%

We then compute the prompt $J/\psi$ production rates at the
$\sqrt{s}=13$~TeV LHC by convolving with the parton production cross sections. 
We follow the method in ref.~\cite{Bodwin:2015iua}, 
resumming both threshold and DGLAP logarithms in the FFs, 
and convolving them with the NLO parton cross sections taken from 
ref.~\cite{Aversa:1988vb}.
We then compare the results with ATLAS 
measurements~\cite{ATLAS:2023qnh} in Fig.~\ref{fig:ATLAS}.
Remarkably, the resummed results are in good agreement with large-$p_T$ data, 
while the fixed-order results at NLO accuracy fail to describe measurement 
as it falls below data and turns negative. 
This shows that resummation of threshold logarithms is imperative in 
describing heavy quarkonium production rates at very large transverse
momentum.

%==============================================================================
\section{Summary and conclusion}
\label{sec:summary}
%==============================================================================

In our recent work in ref.~\cite{Chung:2024jfk} we computed, for the first
time, the complete leading threshold double logarithms that appear in the
nonrelativistic QCD factorization description of the inclusive production rates 
of $J/\psi$, $\psi(2S)$, and $\chi_{cJ}$. 
The resummation of threshold double logarithms that was obtained in 
ref.~\cite{Chung:2024jfk} substantially improves the NRQCD description of
charmonium production rates at large transverse momentum, and resolves the
catastrophic failure of fixed-order perturbation theory where large-$p_T$ cross
sections can turn unphysically negative.

The analysis of threshold logarithms in ref.~\cite{Chung:2024jfk}, 
along with preceding studies in~\cite{Chen:2021hzo, Chung:2023ext, 
Chen:2023gsu}, implies that the negative cross section problem 
is caused by an arbitrary truncation of the perturbation series,
which lead to a failure to acknowledge large radiative corrections. 
This is especially dangerous in quarkonium physics, 
where important theoretical issues can be masked by lack of knowledge in 
nonperturbative matrix elements. 
Despite this, much of the heavy quarkonium production phenomenology is still
strictly based on fixed-order calculations. 
Even the DGLAP evolution, which has been known for nearly half a century, 
has generally not been adopted in calculations of quarkonium production rates, 
outside a few select publications~\cite{Bodwin:2014gia, Bodwin:2015iua, 
Lee:2022anw}. As has been revealed by the resolution of the negative cross 
section problem through threshold resummation, heavy quarkonium production 
processes are just as vulnerable as any other QCD process to the quirks of
perturbative QCD, which are more than often underestimated by attempts to guess
uncertainties from scale variations of fixed-order results. 
The potential for heavy quarkonia to be used as probes of QCD in collider
experiments compels us to take greater care and attention in making theoretical
predictions. 

The formalism that we developed for resummation of threshold logarithms in
ref.~\cite{Chung:2024jfk} also allows us to improve the accuracy of resummation 
through calculation of the soft functions to higher orders
in $\alpha_s$ and $\epsilon$, which can then be used to resum logarithms 
beyond leading double logarithmic level. 
We can expect that the threshold-resummed result will be important in more
differential observables such as the distribution of quarkonium in 
jet~\cite{Baumgart:2014upa, LHCb:2017llq, Kang:2017yde, Bain:2017wvk, 
CMS:2019ebt}. 
It would also be interesting to investigate threshold singularities in NLP
contributions, which become important at low $p_T$.

\begin{acknowledgments}
We thank Geoffrey Bodwin for helpful discussions. 
The work of H.~S.~C. and J.~L. is supported by  the National Research
Foundation of Korea (NRF) Grant funded by the Korea government (MSIT) under
Contract No. NRF2020R1A2C3009918. 
\end{acknowledgments}

\bibliographystyle{JHEP}
\bibliography{thresholdproc.bib}

\end{document}